\documentclass[osajnl,twocolumn,showpacs,superscriptaddress,10pt]{revtex4-1}
\usepackage{epsfig}
\usepackage{amsmath}
\usepackage{amssymb}
\usepackage{graphicx}
\usepackage{indentfirst}
\usepackage[tight]{subfigure}
\usepackage[latin1]{inputenc}

\begin{document}

\title{Coherence properties of coupled optomechanical cavities}
\author{T. Figueiredo Roque}
\email{tfroque@ifi.unicamp.br}
\author{A. Vidiella-Barranco}
\email{vidiella@ifi.unicamp.br}
\affiliation{Instituto de Física Gleb Wataghin, Universidade Estadual de Campinas, 13083-859 Campinas, São Paulo, Brazil}
\date{\today}

\begin{abstract}
In this work we investigate an optomechanical system consisting of two cavities coupled to the same mechanical resonator. We consider each cavity being weakly pumped as well as a small tunneling rate between the cavities. In such conditions, the system can be studied via quantum Langevin equations and the steady state solution can be found perturbatively. In order to ensure that the approximations and methods used to study the system are suitable, the analytical results were compared to numerical simulations. We study the statistical properties of the cavity radiation fields and we show that depending on the values of the parameters of the system, it is possible to modify the spectrum of the cavities and even enhance the sub-Poissonian character of the cavity field.
\end{abstract}
\maketitle

\section*{1. INTRODUCTION}

The field of optomechanics deals with systems in which mechanical and optical degrees of freedom interact with each other. Such systems were first considered in the 70's \cite{braginsky, braginsky2, braginsky3}, in the context of interferometric measurements of gravitational waves, and were extensively studied in the 90's because of their applications in the field of quantum optics \cite{fabre, mancini2, jacobs, pinard}. In the last years however, we have seen a significant increase of interest in optomechanics, especially stimulated by recent experimental achievements, like sideband cooling of the mechanical resonator \cite{gigan, schliesser, thompson, teufel, rocheleau,teufel3}, normal-mode splitting \cite{groblacher, teufel2} and optomechanically induced transparency \cite{teufel2, weis, safavi}. Moreover, there is also the possibility of using optomechanical systems in a variety of technological and scientific applications, like building more sensitive force sensors \cite{marquardt}, applications in quantum information processing \cite{marquardt} and the possibility of bringing quantum phenomena to the macroscopic world \cite{mancini, vitali}. 

Until quite recently, all the experimental realizations of optomechanical systems had a small optomechanical coupling. Consequently, most of the works in optomechanics dealt with the regime of strong pumping (also called linear regime). In this regime the optomechanical coupling parameter $g$ is enhanced by a factor of $\sqrt{n}$, where $n$ is the number of photons of the optical cavity, enabling the observation of the effects of the optomechanical coupling in the laboratory. However, a different optomechanical regime has been considered recently, in which a single photon could interact strongly with the mechanical oscillator. In such a regime the nonlinear nature of the optomechanical coupling could be better explored, giving rise to new possibilities, like the generation of Schroedinger cat states \cite{mancini3, bose}, sub-Poissonian properties of the cavity field \cite{rabl, kronwald, stannigel, qiu}, photon induced tunneling \cite{xu}, non-Gaussian steady states for the mechanical oscillator \cite{nunnenkamp} and enhancement of photon/phonon nonlinearities \cite{stannigel, ludwig}. Experimental realizations in this regime have been done only in cold atomic gases\cite{gupta, brennecke}, but significant progress is being made in this issue.

In this work we consider a system composed by two optical cavities coupled to the same mechanical resonator in the single-photon strong coupling regime. The two cavities are weakly pumped and photons can tunnel from one cavity to the other. We show that the sub-Poissonian properties of the cavity field, predicted by Rabl \cite{rabl}, can be enhanced by choosing appropriately the values of some parameters easily controlled in the laboratory. The results we have found open the possibility of using optomechanical systems as single photon sources. 

Our paper is organized as follows: in section II we quickly review the single cavity system and present our model and its solution; in section III we discuss our results, and in section IV we summarize our conclusions.
    
\section*{2. THE MODEL AND ITS SOLUTION}
\begin{figure}
	\centering	
	\includegraphics[height=4cm]{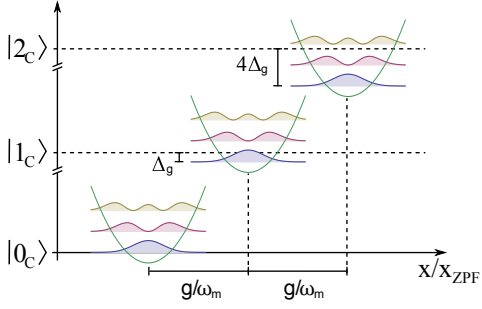}	
	\caption{Energy level diagram of Hamiltonian (\ref{hs1}) for $E=0$. If the cavity is in the $| n_c \rangle$ state, the radiation pressure force displaces the mechanical oscillator by $n_c g/\omega_m$ and the eigenenergies of the cavity are lowered by $\Delta_g n_c^2$. }
	\label{soms}	
\end{figure}

Before considering the system proposed in this paper, it would be enlightening to quickly review the single cavity system studied by Rabl\cite{rabl} and Nunnenkamp et al.\cite{nunnenkamp}. In this system we have an optomechanical system weakly pumped by a coherent field. The Hamiltonian of this system (in a rotating reference frame) is,
\begin{equation}
	\hat{H} = \Delta \hat{a}^{\dagger} \hat{a} + g \hat{a}^{\dagger} \hat{a}(\hat{b} + \hat{b}^{\dagger}) + \omega_m \hat{b}^{\dagger} \hat{b} + i E(\hat{a}^{\dagger} - \hat{a}),
	\label{hs1}
\end{equation} 
where $\Delta = \omega_c - \omega_L$, $\omega_c$ is the frequency of the cavity and $\omega_L$ is frequency of the pumping field. The weak pumping condition means that $E\ll\kappa$, where $\kappa$ is the decay rate of the optical cavity. In the absence of pumping($E=0$) the Hamiltonian (\ref{hs1}) is diagonalized by the polaron transformation,
\begin{equation}
	\hat{U} = \exp\left[g \hat{a}^{\dagger} \hat{a} (\hat{b}^{\dagger} - \hat{b})/\omega_m \right].
\end{equation} 
In the presence of pumping, however, the transformed Hamiltonian takes the following form,
\begin{multline}
	\hat{H}^{\prime} = \hat{U} \hat{H} \hat{U}^{\dagger} = \Delta \hat{a}^{\dagger} \hat{a} - \Delta_g (\hat{a}^{\dagger} \hat{a})^2 + \omega_m \hat{b}^{\dagger} \hat{b}\\ + i E(\hat{a}^{\dagger} e^{-i \hat{P}} - \hat{a} e^{i \hat{P}}),
	\label{hs2}
\end{multline} 
where $\Delta_g = g^2/\omega_m$ and $\hat{P} = i g (\hat{b}^{\dagger} - \hat{b})/\omega_m$. Although we still have a non-linear interaction term between the optical field and the mechanical oscillator, significant simplifications related to the dynamics of the optical field can be done if the cavity is weakly pumped. Neglecting the pumping term for a while, the polaron transformation makes evident that the optomechanical interaction displaces the mechanical oscillator from its equilibrium position by an amount of $g n_c/\omega_m$(where $n_c$ is the number of photons in the cavity) and consequently lowers the energy of the $n_c$-photon state by $n_c^2 \Delta_g$, as can be seen in fig.(\ref{soms}). Because of the Kerr non-linearity induced by the optomechanical coupling, we can observe that if the driving field is resonant with the the transition from the vacuum state to the 1-photon state, it is not resonant with the transition from 1-photon state to the 2-photon state. This phenomenon gives rise to the sub-Poissonian character of the radiation found in \cite{rabl}.  

\begin{figure}
	\centering	
	\subfigure[]{
		\includegraphics[height=2.9cm]{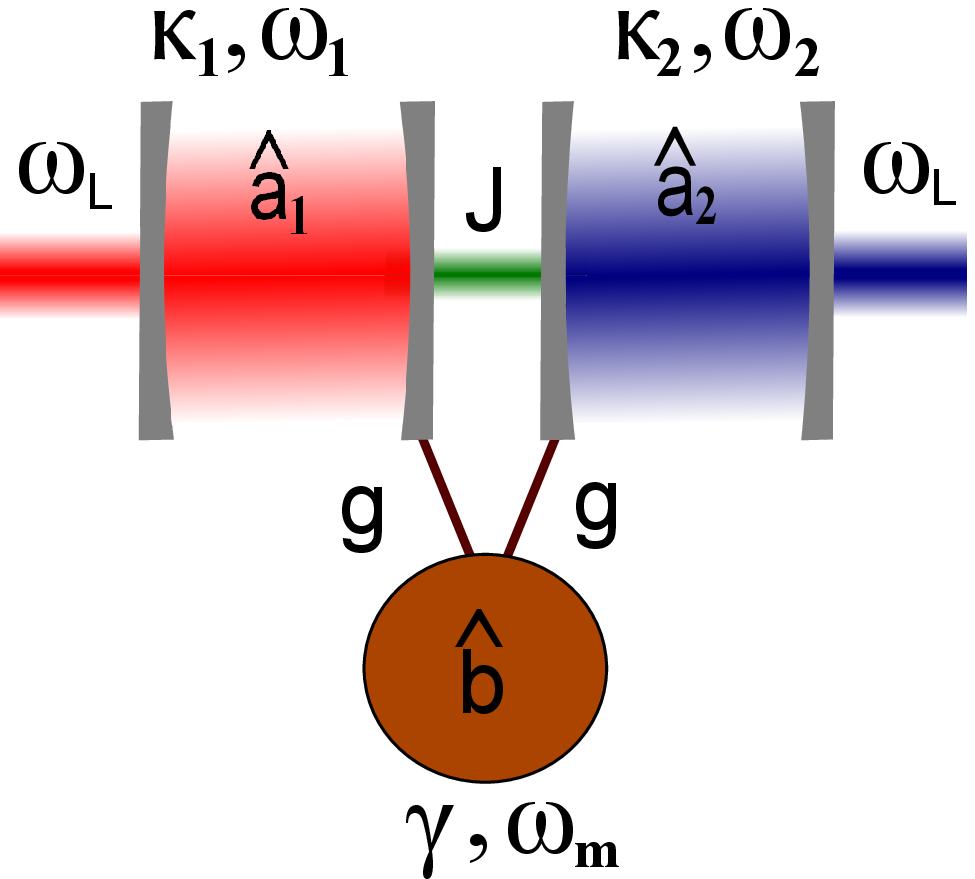}
		\label{oms}
	}
	\subfigure[]{
		\includegraphics[height=5.7cm]{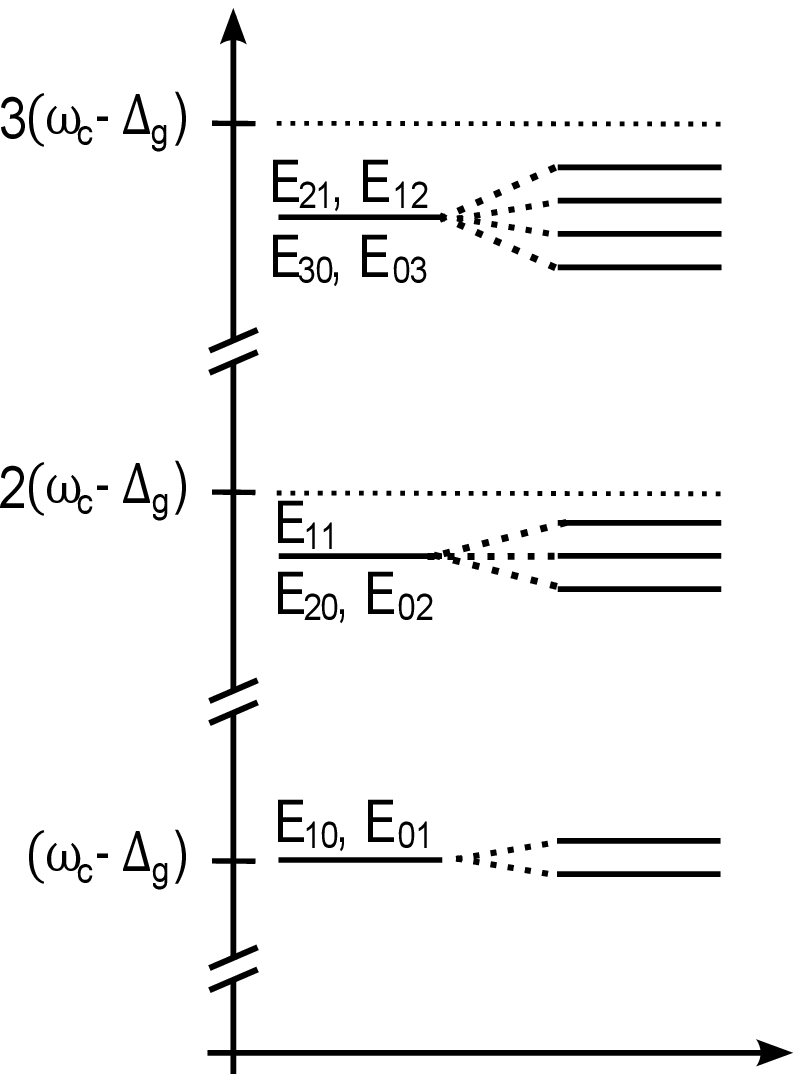}
		\label{energyl}
	}
	\caption{(a)In the optomechanical system considered here we have two optical cavities optomechanically coupled to the same mechanical oscillator with the same optomechanical coupling parameter $g$. The cavities are also tunnel-coupled, with tunneling amplitude $J$, and are weakly pumped by lasers with the same frequency $\omega_L$. The colors of the cavities do not have any relationship with the detuning of the cavities. (b) Energy level diagram of Hamiltonian (\ref{h3}). We shall call the states for which $n_1 + n_2 = 1$ the first group of states, the states for which $n_1 + n_2 = 2$ the second group of states, and the states for which $n_1 + n_2 = 3$ the third group of states.}
	\label{omsenergyl}	
\end{figure}

Our purpose is to enhance the sub-Poissonian character of the radiation field exploring a similar system, which is composed by two cavities optomechanically coupled to the same mechanical resonator with the same coupling constant $g$. Each cavity is pumped by a coherent field with amplitude $E_i(i=1,2)$ (we assume that both coherent fields have the same frequency $\omega_L$) and photons can tunnel from one cavity to the other, where the tunneling constant is $J$. The Hamiltonian of this system (in a rotating reference frame) is:
\begin{multline}
	\hat{H} = \sum\limits_{j=1,2} \Delta_j \hat{a}_j^{\dagger} \hat{a}_j + i E_j(\hat{a}_j^{\dagger} - \hat{a}_j)  + g \hat{a}_j^{\dagger} \hat{a}_j(\hat{b} + \hat{b}^{\dagger}) \\- J(\hat{a}_1^{\dagger}\hat{a}_2 + \hat{a}_1 \hat{a}_2^{\dagger}) + \omega_m \hat{b}^{\dagger} \hat{b},
	\label{h1}
\end{multline}
\noindent
where $j = 1,2$, $\hat{a}_j$ and $\omega_j$ are, respectively, the annihilation operator and the frequency of the j'th cavity, and $\Delta_j = \omega_j - \omega_L$. $\hat{b}$ is the annihilation operator of the mechanical oscillator. If the pumping fields $E_j$ and the tunneling constant $J$ are zero, the Hamiltonian (\ref{h1}) is diagonalized by the polaron transformation,
\begin{equation}
	\hat{U} = \exp\left[g(\hat{a}_1^{\dagger} \hat{a}_1 + \hat{a}_2^{\dagger} \hat{a}_2) (\hat{b}^{\dagger} - \hat{b})/\omega_m \right].
	\label{pt}
\end{equation} 
Thus, our first approximation will be to consider a weak pumping(i.e., $E_j\ll \kappa_j$, where $\kappa_j$ is the decay rate of the optical cavity) and a small tunneling constant $J$(i.e., $J\ll \kappa_j$), so that we can study our system using a perturbative approach. To understand how this system can be used to achieve a stronger sub-Poissonian character for the optical field, lets first apply the transformation (\ref{pt}) to Hamiltonian (\ref{h1}),
\begin{multline}
	\hat{H}^{\prime} = \sum\limits_{j=1,2} \Delta_j \hat{a}_j^{\dagger} \hat{a}_j - \Delta_g (\hat{a}_j^{\dagger} \hat{a}_j)^2 +  i E_j(\hat{a}_j^{\dagger} e^{-i\hat{P}} - \hat{a}_j e^{i\hat{P}}) \\- 2 \Delta_g \hat{a}_1^{\dagger} \hat{a}_1 \hat{a}_2^{\dagger} \hat{a}_2 - J(\hat{a}_1^{\dagger}\hat{a}_2 + \hat{a}_1 \hat{a}_2^{\dagger}) + \omega_m \hat{b}^{\dagger} \hat{b}.
	\label{h2}
\end{multline}
In Hamiltonian (\ref{h2}) the term inside the summation corresponds to an optical cavity with a Kerr non-linearity and a pumping term, in accordance with the result found for a single cavity optomechanically interacting with a mechanical oscillator. However, the interaction of the cavities with the same mechanical oscillator gives rise to a cross-Kerr term, $- 2 \Delta_g \hat{a}_1^{\dagger} \hat{a}_1 \hat{a}_2^{\dagger} \hat{a}_2$. To understand the role of the cross-Kerr term in the system dynamics, lets suppose $E_j=J=0$ and neglect the mechanical oscillator energy term in Hamiltonian (\ref{h2}). In a non-rotating frame with respect to the laser frequency, we have the following Hamiltonian,
\begin{equation}
	\hat{H}^{\prime} = \sum\limits_j \omega_j \hat{a}_j^{\dagger} \hat{a}_j - \Delta_g (\hat{a}_j^{\dagger} \hat{a}_j)^2 - 2 \Delta_g \hat{a}_1^{\dagger} \hat{a}_1 \hat{a}_2^{\dagger} \hat{a}_2.
	\label{h3}
\end{equation}
The Fock states $|n_1, n_2 \rangle$ are eigenstates of Hamiltonian (\ref{h3}) with eigenenergy $E_{n_1,n_2} = \omega_1 n_1 + \omega_2 n_2 - \Delta_g (n_1 + n_2)^2$, the cross-Kerr term being responsible for lowering the energy of the $|n_1, n_2 \rangle$ state by $2 n_1 n_2$. Lets suppose that cavity 1(2) is being pumped by a coherent field in resonance with the transition from the vacuum state to the $|0,1\rangle$($|1,0\rangle$). In this situation, the cross-Kerr term eventually contributes to increase the detuning between the coherent field and the transition from the $|1,0\rangle$ and $|0,1\rangle$ to higher energy states, and this is expected to enhance the sub-Poissonian character of the optical field. However, as will became clear later, in the weak pumping regime, the effect of the cross-Kerr term is negligible, unless the cavities are coupled by tunneling. Thus, it is necessary to sum the term $H_t = - J(\hat{a}_1^{\dagger}\hat{a}_2 + \hat{a}_1 \hat{a}_2^{\dagger})$ in Hamiltonian (\ref{h3}). To better understand this new situation, lets consider, for simplicity, $\omega_1 = \omega_2$. In this case, the presence of tunneling define new eigenstates and eigenenergies for our system, as represented in fig. \ref{energyl}. The first four states are shown below:
\begin{equation}
	|1, \pm 1 \rangle_J = (|1,0\rangle \pm |0,1\rangle)/\sqrt{2},\quad E_{1,\pm 1} = \omega - \Delta_g \pm J 
	\label{es1}
\end{equation}
\begin{multline}
	|2, \epsilon \rangle_J = |2,0\rangle + \sqrt{2} \epsilon |1,1\rangle + |0,2\rangle,\\ E_{2,\epsilon} = 2 \omega - 4 \Delta_s - 2 \epsilon J,
	\label{es2}
\end{multline}
where $\epsilon = 0, \pm 1$. As we can see, the tunneling rises the energy of some levels, decreasing the detuning between the energy of the incoming photons and the energy necessary to add a photon. However, if $J$ is small enough, it is still expected to observe sub-Poissonian statistics in the radiation. For bigger values of $J$, some energy levels will be very close to resonance with the incoming photons, so it is expected that the sub-Poissonian character of the radiation should then be reduced.

We will now develop the perturbative treatment of the system under study in order to analyze it more rigorously. As we have seen before, after making the polaron transformation (\ref{pt}) in Hamiltonian (\ref{h1}), we obtain Hamiltonian (\ref{h2}). Supposing that the system is coupled to its environment, a suitable way of studying the optical field is via quantum Langevin equations (QLE). In the transformed reference frame, the QLE for $\hat{a}_j$ is,
\begin{multline}
	\frac{d}{dt}\hat{a}_j = -(\kappa_j/2 + i \Delta_j) \hat{a} - i \Delta_g \bigl(\hat{a}_j + 2 \hat{a}^{\dagger}_j \hat{a}^2_j + 2 \hat{a}_j \hat{a}^{\dagger}_p \hat{a}_p\bigr) \\+ i J \hat{a}_p + e^{-i \hat{P}}\bigr(E_j - \sqrt{\kappa_j} \hat{\xi}_{j}\bigr), 
	\label{qle}
\end{multline}
where $j, p = 1,2$ and $p \neq j$. $\kappa_j$ is the decay rate of the j'th cavity and $\hat{\xi}_{j}(t)$ is the input noise on the j'th cavity, which satisfy the relations $\langle \hat{\xi}_{j}(t) \hat{\xi}_{j}^{\dagger}(t') \rangle = \delta(t - t')$ and $\langle \hat{\xi}_{j}^{\dagger}(t) \hat{\xi}_{j}(t') \rangle = 0$. The QLE for $\hat{b}$ is,
\begin{multline}
	\frac{d}{dt} \hat{b} = -(\gamma + i \omega_m) \hat{b} + \sum\limits_{j=1,2}\frac{E_j g}{\omega_m} \bigl(\hat{a}^{\dagger}_j e^{-i \hat{P}} + \hat{a}_j e^{i \hat{P}}\bigr) \\+ \sqrt{\gamma} \hat{\chi}
	\label{qle2}
\end{multline}
where $\gamma$ is the decay rate and $\hat{\chi}$ is the input noise on the mechanical oscillator. The latter satisfy the relations $\langle \hat{\chi}(t) \hat{\chi}^{\dagger}(t') \rangle = (\overline{n}+1)\delta(t - t')$ and $\langle \hat{\chi}^{\dagger}(t) \hat{\chi}(t') \rangle = \overline{n} \delta(t - t')$, where $\overline{n} = [\exp(\hbar \omega_m/K_B T) - 1]^{-1}$ is the mean thermal excitation number.
The weak pumping regime allow us to make some simplifications in eq. (\ref{qle}). As we are interested in calculating normally ordered averages, and given that in the steady state $\hat{a}_j \propto E_j + \hat{\xi}_{j}$, the non-linear terms in eq. (\ref{qle}) can be neglected, as they give rise only to higher order terms in the pumping fields. This implies that the cross-Kerr term is irrelevant to the dynamics of $a_j(t)$ and, if $J=0$, the cavities would behave as if there was no other cavity coupled to the mechanical oscillator, i.e., the situation explored in \cite{rabl} and \cite{nunnenkamp}. Following the same argument, given that the effect of the optical field in the mechanical oscillator is of second order in $E_j$, we can approximate eq. (\ref{qle2}) by the equation of motion of a simple dissipative harmonic oscillator,
\begin{equation}
	\frac{d}{dt}\hat{b} = -(\gamma + i \omega_m) \hat{b} + \sqrt{\gamma} \hat{\chi}.
\end{equation} 
With the above approximations, eq. (\ref{qle}) can be solved analytically. However, to simplify even more our analysis, we will suppose that $\gamma$ is very small compared to $\kappa_{j}$, so that we can approximate $\hat{b}(t)$ by its free evolution and $\hat{P}(t)$ is given by,
\begin{equation}
	\hat{P}(t) = i g \bigl[\hat{b}^{\dagger}(0) e^{i \omega_m t} - \hat{b}(0) e^{-i \omega_m t}\bigr]/\omega_m. 
\end{equation}
Thus, $\hat{P}(t)$ is determined by its free evolution from an initial state, which we assume to be a thermal state at zero temperature. Under these circumstances, the steady state solution for the eq. (\ref{qle}), transformed back to the original frame, is
\begin{multline}
	\hat{a}_j(t) = \int\limits_{-\infty}^t d\tau_0 \exp\bigl[-(\kappa_j + i \tilde{\Delta}_j)(t - \tau_0)\bigr] e^{i\hat{P}(t)} \times \\e^{-i\hat{P}(\tau_0)} \bigl[i J \hat{a}_p(\tau_0) + \hat{a}_{j, in}(\tau_0)\bigr],
	\label{sss}
\end{multline}
where $\tilde{\Delta}_j = \Delta_j - \Delta_g$ and $\hat{a}_{j, in}(t) = E_j - \sqrt{\kappa_j} \hat{\xi}_{j}(t)$. But eq. (\ref{sss}) is still a function of $\hat{a}_p(t)$. Inserting the equation for $\hat{a}_p(t)$ in the equation for $\hat{a}_j(t)$ we obtain an integral equation for $a_j(t)$ whose solution can be obtained recursively, 
\begin{equation}
	\hat{a}_j(t) = \sum\limits_{n=0}^{\infty} \hat{a}_{j,n}(t) J^n,
	\label{sa}
\end{equation}
where,
\begin{multline}
	\hat{a}_{j,n}(t) = \int\limits_{-\infty}^{t} d\tau_0 \int\limits_{-\infty}^{\tau_0} d\tau_1 \ldots \int\limits_{-\infty}^{\tau_{n-1}} d\tau_n i^n \exp\Bigl[-\Bigl(\frac{\kappa_j}{2} + i \tilde{\Delta}_j\Bigr)\times\\(t - \tau_0)-\left(\frac{\kappa_p}{2} + i \tilde{\Delta}_p\right)(\tau_0 - \tau_1) - \ldots -\left(\frac{\kappa_r}{2} + i \tilde{\Delta}_r\right)\times\\ (\tau_{n-1} - \tau_n)\Bigr] e^{i \hat{P}(t)} e^{-i \hat{P}(\tau_n)} \hat{a}_{r, in}(\tau_n),
	\label{csa}
\end{multline}
with $r = j$ if $n$ is even and $r = p$ if $n$ is odd. It is important to observe that in the weak pumping regime with $J=0$, the two cavities behave as independent optomechanical systems. This is expected as the interaction of the two cavities via mechanical oscillator is of second order in the pumping fields. From eqs. (\ref{sa}) and (\ref{csa}) the excitation spectrum of the j'th cavity,
\begin{equation}
	S_j = \frac{\kappa_j^2}{4 E_j^2} \lim\limits_{t \to \infty} \langle \hat{a}_j^{\dagger}(t) \hat{a}_j(t) \rangle,
\end{equation}
can be readily calculated. However the result will depend on the second order correlation function of the resonator $\langle e^{i \hat{P}(\tau)} e^{-i \hat{P}(\tau^{\prime})} \rangle$ and the best way to compute the spectrum is by expanding the resonator correlation function in a series of exponentials \cite{rabl}.

To study the second order coherence properties of the radiation we must also find the steady state solution of $\hat{a}_j^2$. The QLE for $\hat{a}_j^2$ is, 
\begin{multline}
	\frac{d}{dt} \hat{a}_j^2 = -(\kappa_j + 2 i \Delta_j)\hat{a}_j^2 - 4i \Delta_g\bigl(\hat{a}^2_j + \hat{a}^{\dagger}_j \hat{a}^3_j + \hat{a}^2_j \hat{a}^{\dagger}_p \hat{a}_p\bigr) \\+ 2i J \hat{a}_1 \hat{a}_2 + 2 e^{-i \hat{P}} \hat{a}_j \hat{a}_{j, in}.
	\label{qle3}
\end{multline}
\noindent
By the same arguments used to simplify eq. (\ref{qle}), we simplify eq. (\ref{qle3}) by neglecting its non-linear terms. At first sight, one would think that the cross-Kerr term has little relevance in eq. (\ref{qle3}), as the term  $4 \Delta_g \hat{a}^2_j \hat{a}^{\dagger}_p \hat{a}_p$ is going to be neglected. However, this is not true, as we are going to see below. The steady state solution of eq. (\ref{qle3}) is,
\begin{multline}
	\hat{a}_j^2(t) = \int_{-\infty}^{t} d\tau_0 \exp\bigl[-(\kappa_j + 2i \tilde{\Delta}_j - 2i \Delta_g) (t-\tau_0) \bigr] e^{2i \hat{P}(t)}\\ \times e^{2i \hat{P}(\tau_0)} \bigl\{ 2 \hat{a}_j(\tau_0) \hat{a}_{j, in}(\tau_0) + 2i J \hat{a}_1(\tau_0) \hat{a}_2(\tau_0) \bigr\}.
	\label{sss2}
\end{multline}
However eq. (\ref{sss2}) still depends on $\hat{a}_1(t) \hat{a}_2(t)$, which satisfy the following QLE,
\begin{multline}
	\frac{d}{dt} \bigl(\hat{a}_1\hat{a}_2\bigr) = -\Bigl(\frac{\kappa_1}{2} + \frac{\kappa_2}{2} + i \Delta_1 + i \Delta_2 \Bigr) \hat{a}_1\hat{a}_2 - 4i \Delta_g \bigl(\hat{a}_1 \hat{a}_2 \\+ \hat{a}_1^{\dagger}\hat{a}_1^2 \hat{a}_2 + \hat{a}_2^{\dagger}\hat{a}_2^2 \hat{a}_1 \bigr)+ iJ \bigl(\hat{a}_1^2 + \hat{a}_2^2\bigr) + e^{-i \hat{P}} \hat{a}_2 \hat{a}_{1, in} \\+ e^{-i \hat{P}} \hat{a}_1 \hat{a}_{2, in}.
	\label{qle4}
\end{multline}
By the same arguments used to simplify eqs. (\ref{qle}) and (\ref{qle3}), we neglect the non-linear terms in eq. (\ref{qle4}). Nonetheless, the resulting equation has explicit contributions from the cross-Kerr term, given by $-2 \Delta_g \hat{a}_1 \hat{a}_2$. Therefore, eq. (\ref{qle3}) has contributions of the cross-Kerr term due to its dependence on $\hat{a}_1 \hat{a}_2$. The importance of having tunneling between the two cavities now becomes clear, since the dependence of eq. (\ref{qle3}) on $\hat{a}_1 \hat{a}_2$ is proportional to $J$. Again, for $J=0$ we have the same results obtained in \cite{rabl}. Proceeding in the same way as before, the steady state solution of eq. (\ref{qle4}) is,
\begin{multline}
	\hat{a}_1(t)\hat{a}_2(t) = \int\limits_{-\infty}^t d\tau \exp\Bigl[-\Bigl(\frac{\kappa_1}{2} + \frac{\kappa_2}{2} + i \tilde{\Delta}_1 + i \tilde{\Delta}_2 -\\ 2i \Delta_g\Bigr)(t-\tau)\Bigr] e^{2i \hat{P}(t)} e^{-2i \hat{P}(\tau)}\Bigl\{ iJ \bigl[\hat{a}_1^2(\tau) + \hat{a}_2^2(\tau) \bigr] +\\ \hat{a}_{2,in}(\tau) \hat{a}_1(\tau) +  \hat{a}_{1,in}(\tau) \hat{a}_2(\tau)\Bigr\}.
	\label{sss3}
\end{multline}

With eqs. (\ref{sss2}) and (\ref{sss3}) we can obtain the solution for $a_j^2(t)$ recursively. The resulting expression, however, is too cumbersome to be written here. Finally, the second order coherence function of the j'th cavity,
\begin{equation}
	g_{j}^{(2)}(0) = \lim_{t \to \infty} \langle (\hat{a}_j^{\dagger})^2(t) \hat{a}_j^2(t) \rangle / \langle \hat{a}_j^{\dagger}(t) \hat{a}_j(t) \rangle^2,
\end{equation}
can be readily calculated. The result will depend on the correlation function of the resonator, $\langle e^{i \hat{P}(\tau)} e^{i \hat{P}(\tau^{\prime})} e^{-i \hat{P}(\tau^{\prime \prime})} e^{i \hat{P}(\tau^{\prime\prime\prime})} \rangle$. Again, the best way to compute $g_{j}^{(2)}(0)$ is by expanding the correlation function in a series of exponentials. The statistical properties of the radiation are said to be sub-Poissonian if $g^{(2)}(0)<1$, Poissonian if $g^{(2)}(0) = 1$ and super-Poissonian if $g^{(2)}(0) > 1$.

It is also possible to study the system considered here using the following master equation,
\begin{equation}
	\frac{d}{dt} \hat{\rho} = \frac{i}{\hbar} [\hat{\rho}, \hat{H}] + \kappa_1 \mathcal{D}[\hat{a}_1] + \kappa_2 \mathcal{D}[\hat{a}_2] + \gamma (\overline{n}+1) \mathcal{D}[\hat{b}] + \gamma \overline{n} \mathcal{D}[\hat{b}^{\dagger}],
	\label{me}
\end{equation}  
where $\mathcal{D}[\hat{c}] = \hat{c} \hat{\rho} \hat{c}^{\dagger} - \hat{c}^{\dagger} \hat{c} \hat{\rho}/2 - \hat{\rho} \hat{c}^{\dagger} \hat{c}/2$ and $\overline{n} = [\exp(\hbar \omega_m/K_B T) - 1]^{-1}$. The steady state solution of this master equation can be hard to find analytically, but can be easily found employing numerical techniques based on the inverse power method. In this work the steady state solution of eq. (\ref{me}) was found numerically using the QOToolbox \cite{tan}.

\section*{3. RESULTS}
\begin{figure}[!t]
	\centering		
	\subfigure[]{
			\includegraphics[height=3.6cm]{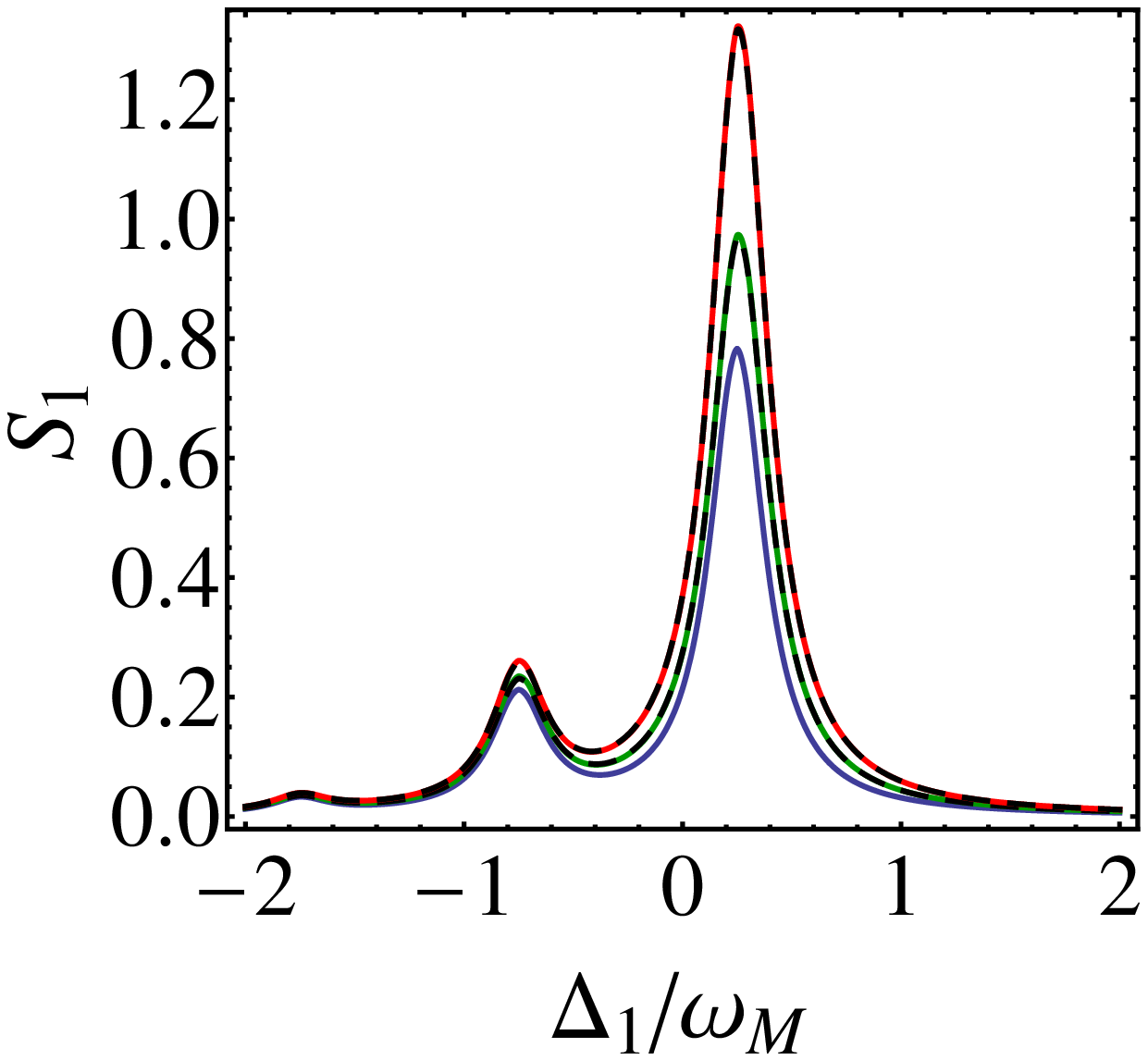}
			\label{fig3a}
	}\hspace{-0.3cm}
	\subfigure[]{
		\includegraphics[height=3.7cm]{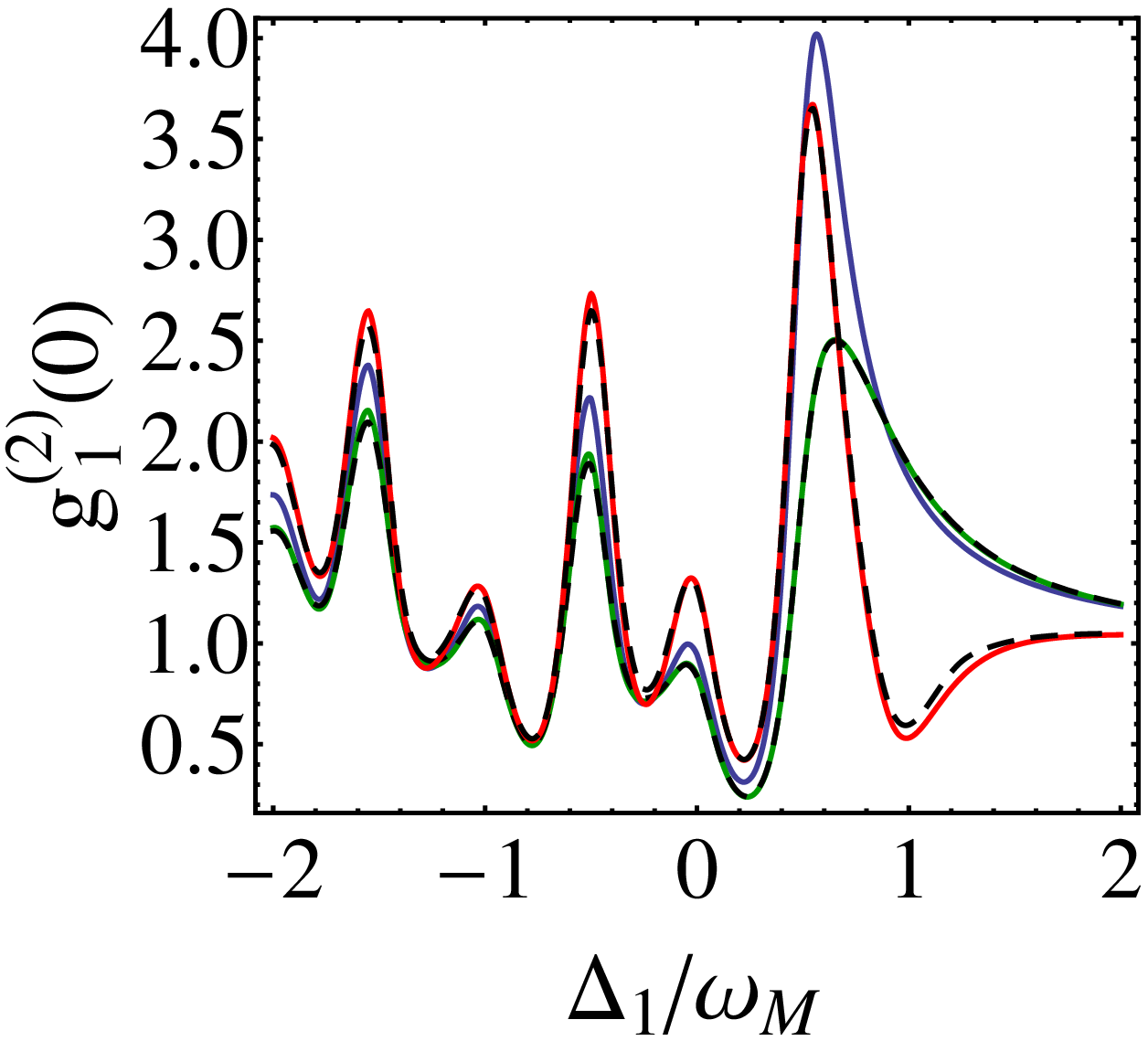}
		\label{fig3b}
	}
	\caption{(a)Spectrum of the cavity 1 against $\Delta_1$. In the blue curve we have $J=0$, in the green curve $J = 0.05 \omega_m$, $\Delta_2 = 0.4\omega_m$ and $E_2 = 0.001\omega_m$ and in the red curve $J = 0.05 \omega_m$, $\Delta_2 = 0.4\omega_m$ and $E_2 = 0.002\omega_m$. The dashed curves are numeric simulations. (b)$g_{1}^{(2)}(0)$ against $\Delta_1$. In the blue curve we have $J=0$, in the green curve $J = 0.05 \omega_m$, $\Delta_2 = 0.4\omega_m$ and $E_2 = 0.001\omega_m$ and in the red curve $J = 0.05 \omega_m$, $\Delta_2 = 0.1\omega_m$ and $E_2 = 0.001\omega_m$. The dashed curves are numerical simulations. The others parameters, which are the same in both graphics and in all curves, are $g = 0.5 \omega_m$, $E_1 = 0.001 \omega_m$, $\kappa_1 = \kappa_2 = 0.3 \omega_m$ and $\gamma = 0.005\omega_m$.}
	\label{fig2}
\end{figure}
\begin{figure}[!t]
	\centering		
	\subfigure[]{
			\includegraphics[height=3.7cm]{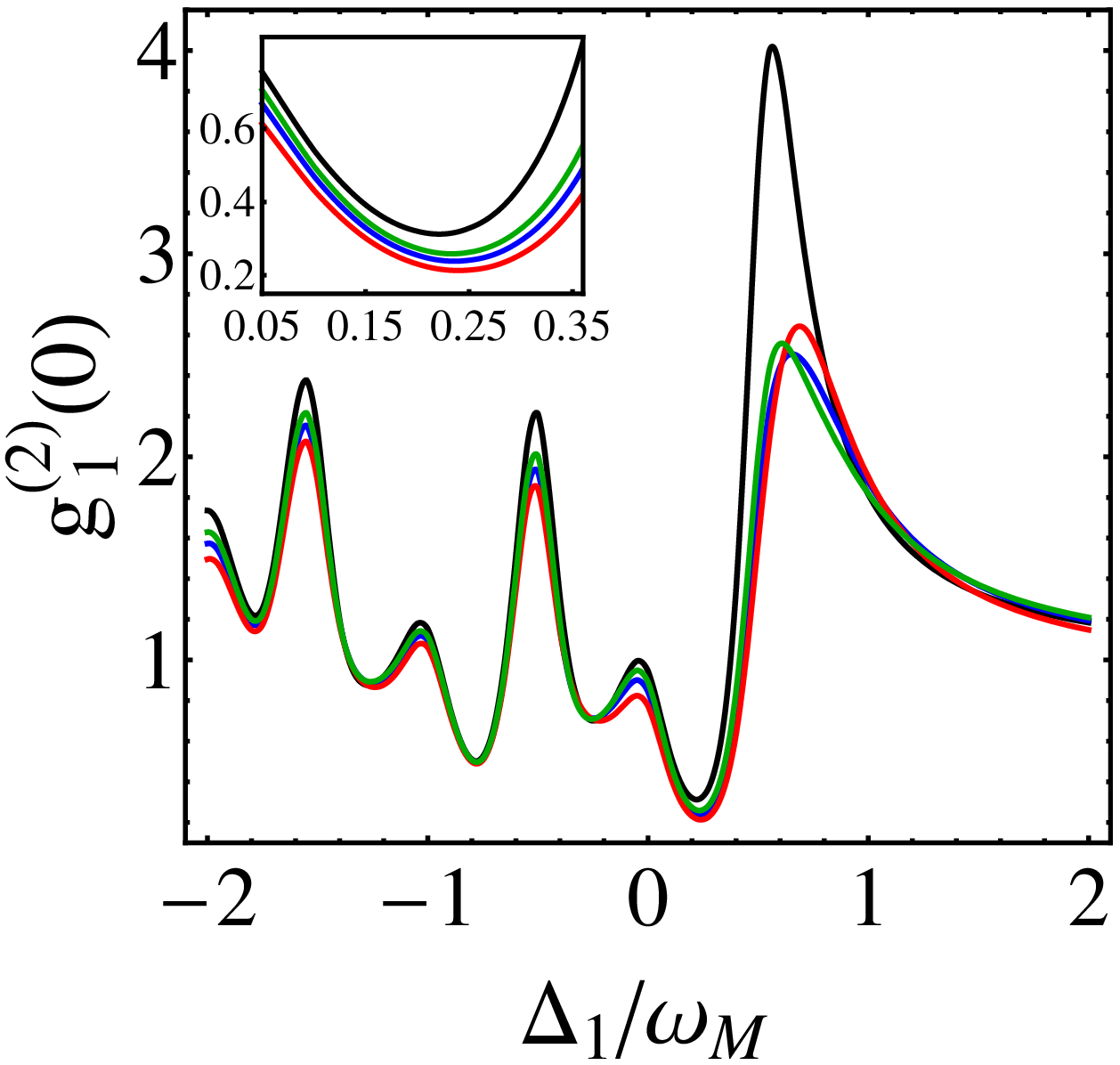}
			\label{fig4a}
	}\hspace{-0.3cm}
	\subfigure[]{
		\includegraphics[height=3.7cm]{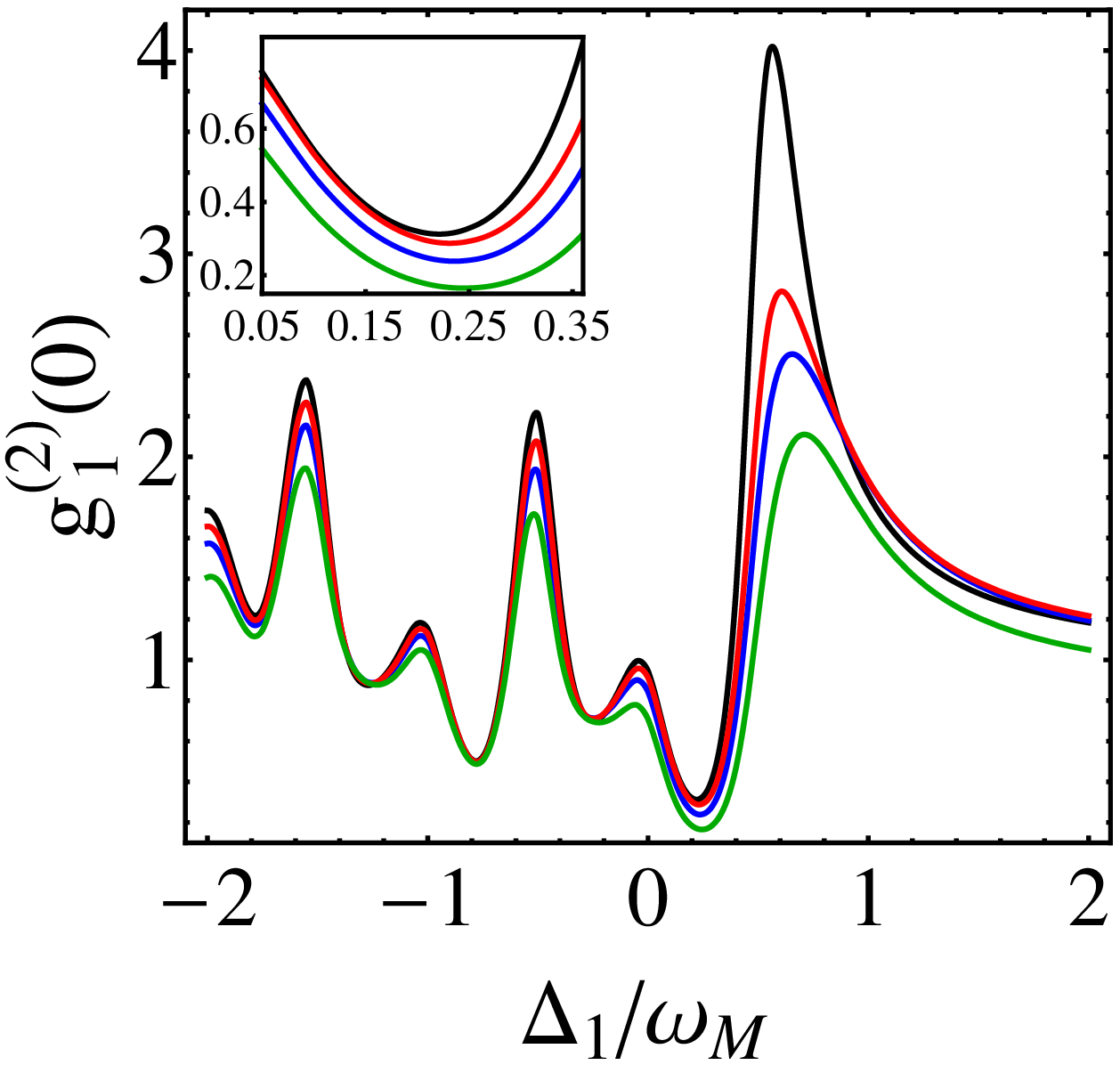}
		\label{fig4b}
	}
	\caption{(a) $g_1^{(2)}(0)$ against $\Delta_1$ for $E_2 = E_1$ and for different values of $\kappa_2$. The black curve is a reference curve with $J=0$, in the blue curve $J=0.05 \omega_m$ and $\kappa_2 = 0.3 \omega_m$, in the red curve $J=0.05 \omega_m$ and $\kappa_2 = 0.2 \omega_m$ and in the green curve $J=0.05 \omega_m$ and $\kappa_2 = 0.4 \omega_m$. (b) $g_1^{(2)}(0)$ against $\Delta_1$ for $\kappa_1 = \kappa_2$ and for different values of $E_2$. The black curve is a reference curve with $J=0$, in the blue curve $J=0.05 \omega_m$ and $E_2 = E_1$, in the red curve $J=0.05 \omega_m$ and $E_2 = E_1/2$ and in the green curve $J=0.05 \omega_m$ and $E_2 = 2 E_1$. In both graphics and in all curves $g=0.5 \omega_m$, $\kappa_1 = 0.3 \omega_m$, $E_1 = 0.001 \omega_m$ and $\Delta_2 = 0.4 \omega_m$.}
	\label{fig4}
\end{figure}
\noindent
As pointed out in the last section, now we can readily compute the excitation spectrum and the second order coherence function $g_{j}^{(2)}(0)$ of the fields. In this section we are going to focus on the properties of cavity 1 and specially on the effects of varying the parameters of cavity 2 and the tunneling constant $J$. In fig.\ref{fig3a} we have the excitation spectrum S of cavity 1 against $\Delta_1$ for different values of $E_2$ and $J$, the continuous curves were obtained using the analytical approach developed here and the dashed curves are numerical simulations. It is interesting to observe that for a single cavity in the weak pumping regime, the excitation spectrum $S$ depends only on the detuning $\Delta$ and on the decay rate of the cavity $\kappa$. However, in our system, due to the possibility of tunneling between the cavities, $S$ depends also on the ratio $E_1/E_2$. Thus, if $J \neq 0$, we are able to modify $S$ just by varying $E_2$ - compare the blue curve $(J = 0)$ with the other curves. The numerical simulations show excellent agreement with the analytical calculations. Fig.\ref{fig3b} shows the second order coherence function $g_{1}^{(2)}(0)$ against $\Delta_1$ for different values of $\Delta_2$. As we see in those plots, $g_{1}^{(2)}(0)$ is very sensitive to variations in $\Delta_2$ but, as we are more interested in the quantum properties of the cavity radiation field, we may take a closer look in the regions were $g_{1}^{(2)}(0)<1$. In fig.\ref{fig3b}, the smallest values for $g_{1}^{(2)}(0)$ for all curves were found for $\Delta_1 \approx 0.25 \omega_m$. This is consistent with the results in \cite{rabl}, where is shown that, in the case of a single optomechanical cavity, there exists a minimum for $g^{(2)}(0)$ at $\Delta = g^2/\omega_m$ (if $\kappa/2 < \omega_m, g$). As we are dealing with a small tunneling constant between the cavities, it is not a surprise that we have found a similar result. However, the interesting fact is that this minimum value for $g_{1}^{(2)}(0)$ can be decreased or increased depending on the value of $\Delta_2$. Our results show that the smallest values for $g_{1}^{(2)}(0)$ for $\Delta_1 = g^2/\omega_m$ are found for $\Delta_2$ slightly larger than $\Delta_1$. This means that the minimum in $g_{1}^{(2)}(0)$ occurs for $\omega_2$ slightly larger than $\omega_1$, since we are assuming that the pumping fields have the same frequency. Using the simplified picture of Hamiltonian (\ref{h3}) and the energy levels of fig. \ref{energyl}, this last fact can be understood as follows. Lets call $|\psi \rangle$ some eigenstate Hamiltonian (\ref{h3}) and $|n, m \rangle$ the state where cavity 1 has $n$ photons and cavity 2 has $m$ photons. In the situation where $\omega_1 = \omega_2$, as the cavities are identical, $|\langle n, m| \psi \rangle|^2 = |\langle m, n| \psi \rangle|^2$. Thus we have equal probability of measuring $n$ photons in cavity 1 and $m$ photons in cavity 2 or $m$ photons in cavity 1 and $n$ photons in cavity 2. But, if $\omega_1 < \omega_2$ we do not have this symmetry anymore, and the probability of measuring more photons in cavity 2 than in cavity 1 is bigger than the probability of measuring the converse. For instance, if $|\psi \rangle$ is the highest energy state of the second group of states, then we have higher probability of measuring the state $|0,2\rangle$ than the state $|2,0\rangle$. Therefore, $g_{1}^{(2)}(0)$ is lowered at expense of increasing the $g_{2}^{(2)}(0)$, as transitions to states with more than one photon in cavity 2 are stimulated. Nonetheless, increasing the difference between $\omega_1$ and $\omega_2$ make the tunneling interaction less effective. This can be seen if go to an interaction picture with respect to both cavities, the tunneling term will become $\hat{H}_t = J (\hat{a}^{\dagger} \hat{b} e^{(\omega_1 - \omega_2) t} + \hat{a} \hat{b}^{\dagger} e^{-(\omega_1 - \omega_2) t})$, which becomes rapidly oscillating if the difference between the frequencies increase. Apart from that, the tunneling between the cavities is crucial to enhance the effects of the cross-Kerr term. Thus, the competition between these two effects determine an optimal value for $\omega_2$ (or $\Delta_2$).
\begin{figure}[!t]
	\centering		
	\subfigure[]{
			\includegraphics[height=3.4cm]{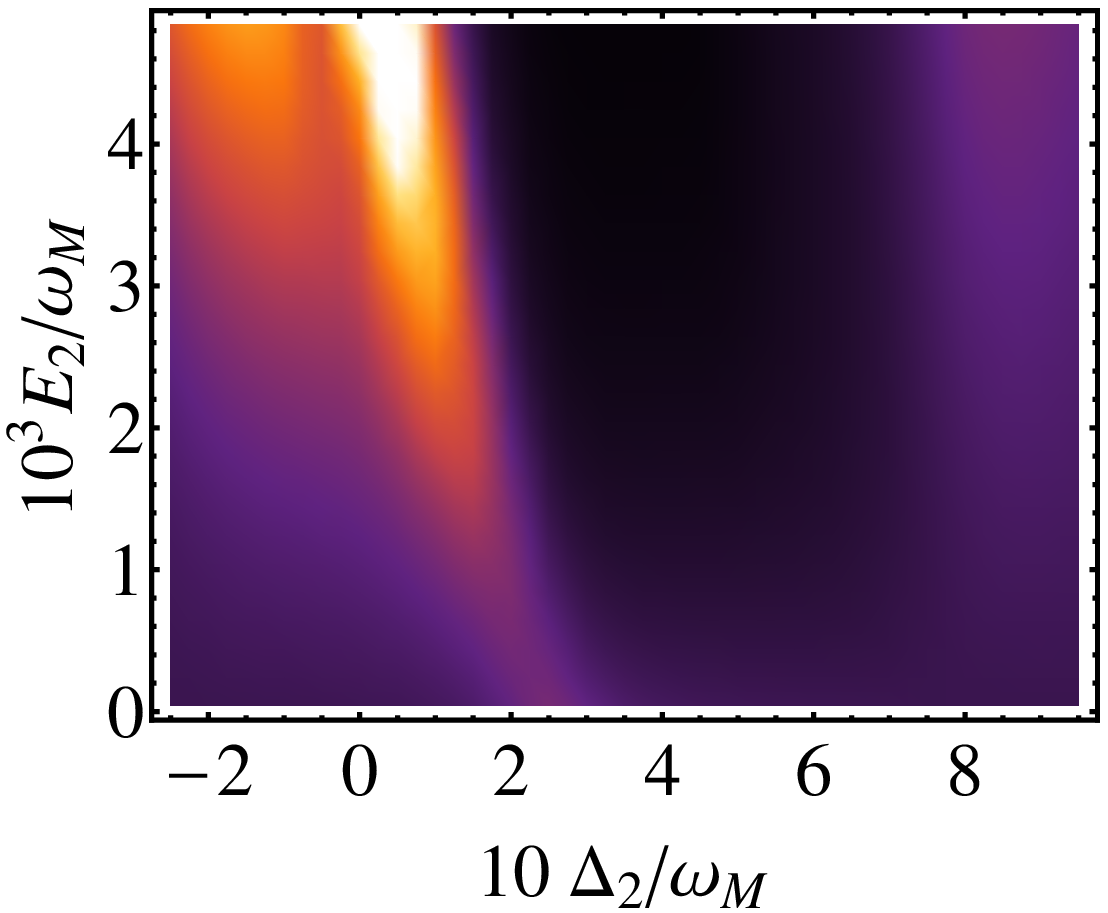}
			\label{fig5a}
		}\hspace{-0.3cm}
	\subfigure[]{
		\includegraphics[height=3.5cm]{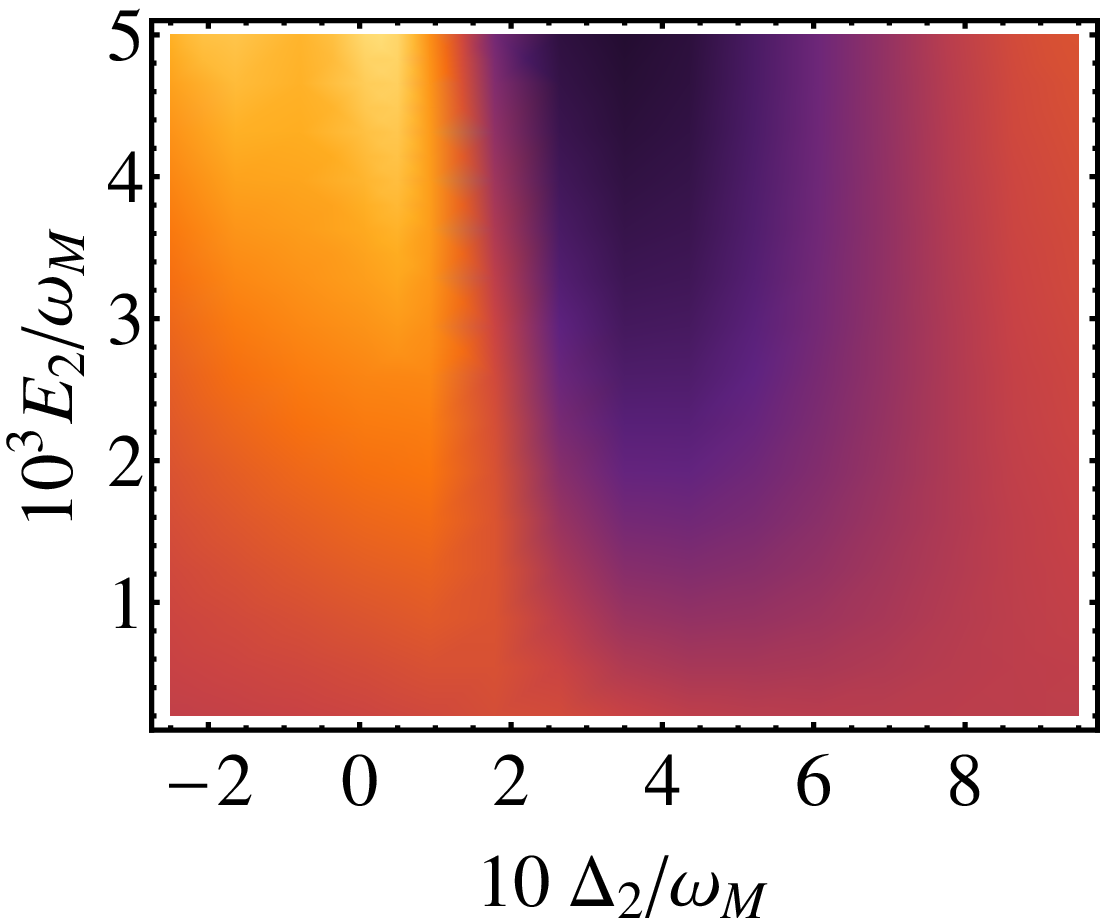}
		\label{fig5b}
	}\hspace{-0.3cm}			
	\subfigure[]{
		\includegraphics[height=3.6cm]{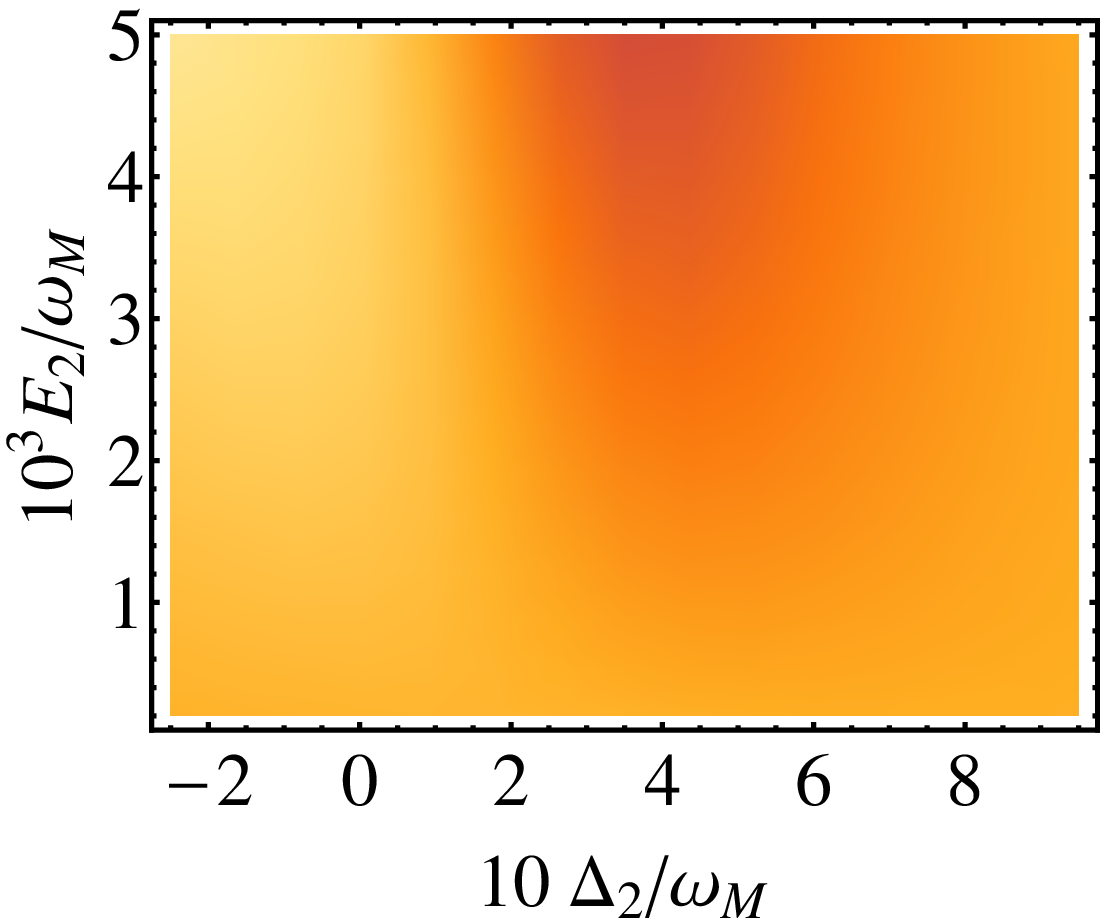}
		\label{fig5c}	
	}\hspace{0.1cm}		
	\subfigure[]{
		\includegraphics[height=3.5cm]{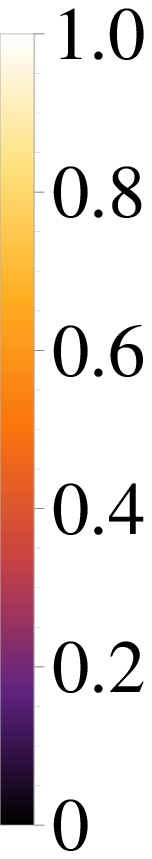}
		\label{fig5d}
	}		
	\caption{The minimum value of $g_{1}^{(2)}(0)$ as a function of $E_2$ and $\Delta_2$ for different values of $\kappa_1$, $\kappa_2$ and $J$. (a)$\kappa_1 = \kappa_2 = 0.15 \omega_m$ and $J = 0.05 \omega_m$. (b)$\kappa_1 = \kappa_2 = 0.3 \omega_m$ and $J = 0.05 \omega_m$. (c) $\kappa_1 = \kappa_2 = 0.6 \omega_m$ and $J = 0.05 \omega_m$. (d) Legend. In all graphics $g = 0.5 \omega_m$ and $E_1 = 0.001 \omega_m$. All the graphics were obtained using the analytical approach developed here.}
	\label{fig5}
\end{figure}
\begin{figure}
	\centering
	\subfigure[]{
		\includegraphics[height=3.7cm]{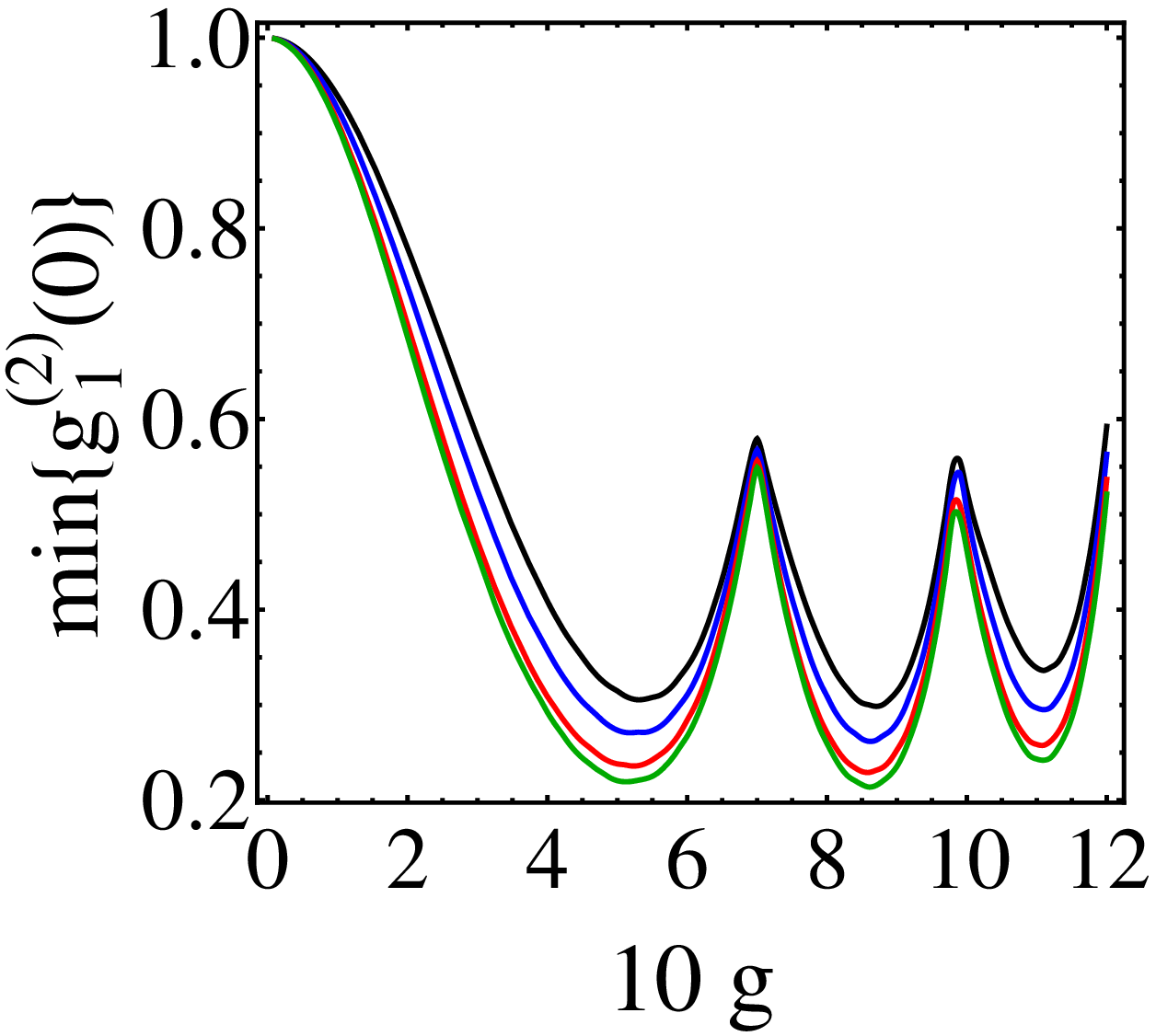}
		\label{fig6a}
	}\hspace{-0.3cm}	
	\subfigure[]{
		\includegraphics[height=3.7cm]{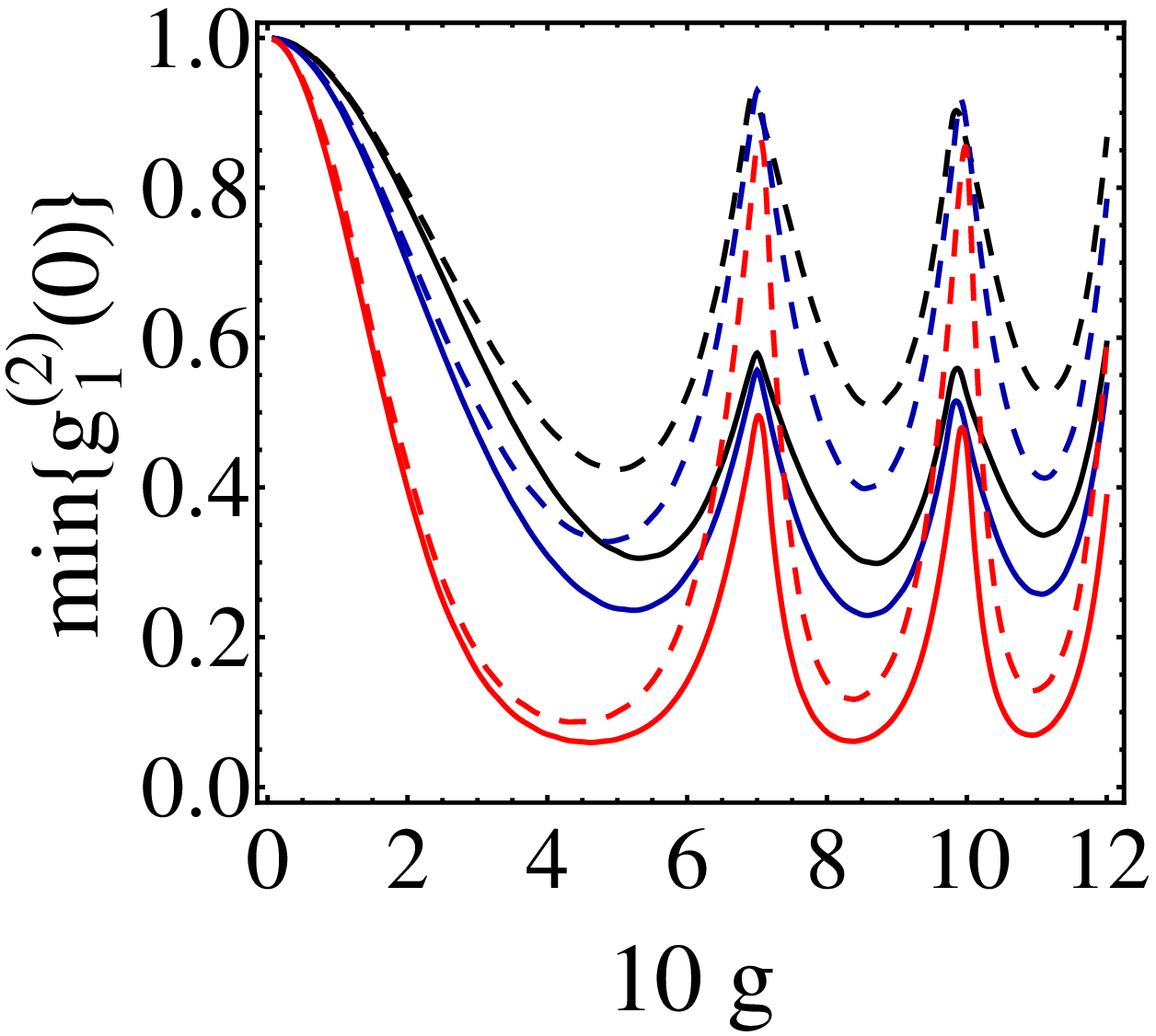}
		\label{fig6b}
	}
	\caption{(a) The minimum value of $g_{1}^{(2)}(0)$ against $g$ for different values of $J$ and $E_1 = E_2 = 0.001 \omega_m$. In the black curve $J=0$, in the blue curve $J=0.02 \omega_m$, in the red curve $J=0.05 \omega_m$ and in the green curve $J = 0.08 \omega_m$. (b) The minimum value of $g_{1}^{(2)}(0)$ against $g$ for different values of $E_2$. The continuous curves were obtained using the analytical approach presented in this paper (mechanical oscillator in contact with a heat bath at zero temperature, $\overline{n}=0$) and the dashed curves were obtained using numerical simulation with the mechanical oscillator in contact with a heat bath with $\overline{n}=1$.	In the black curve and in the black dashed curve we have $J=0$, in the blue curve and in the blue dashed curve we have $J=0.05 \omega_m$ and $E_2 = E_1 = 0.001 \omega_m$, and in the red curve and in the red dashed curve we have $J=0.05 \omega_m$ and $E_2 = 5 E_1 = 0.005 \omega_m$. In the numerical simulations $\gamma = 0.001 \omega_m$.  In both graphics and in all curves $\kappa_1 = \kappa_2 = 0.3 \omega_m$.}
	\label{fig6}	
\end{figure}

In fig.\ref{fig4a} we have $g_{1}^{(2)}(0)$ against $\Delta_1$ for different values of $\kappa_2$. The results show that for smaller values of $\kappa_2$ the system can reach smaller values for $g_{1}^{(2)}(0)$. This is expected as for smaller values of $\kappa_2$, the system is more isolated from the environment and consequently, quantum features, like sub-Poissonian statistics, become more prominent. Another interesting result is that even in the situation where $\kappa_2> \kappa_1$ we obtain smaller values for $g_{1}^{(2)}(0)$ than in the case of a single optomechanical system (or the situation where $J=0$). In fig.\ref{fig4b}, $g_{1}^{(2)}(0)$ is plotted for different values of $E_2$. We note that by increasing the value of $E_2$ we obtain smaller values for $g_{1}^{(2)}(0)$ in the region where $g_{1}^{(2)}(0) < 1$ (the region where the radiation presents a sub-Poissonian character), and that the reduction of $g_{1}^{(2)}(0)$ comes at the expense of increasing $g_{2}^{(2)}(0)$.


In fig.\ref{fig5} we have a color plot of the minimum value of $g_{1}^{(2)}(0)$ against $\Delta_2$ and $E_2$, for different values of $\kappa_1$, $\kappa_2$ and $J$. The plots clearly show the general features pointed out before, such as the fact that the sub-Poissonian character is stronger for $\Delta_2$ slightly larger than $g^2/\omega_m = 0.25$ and for $E_2>E_1$. The results also show that it is possible to reach very small values for $g_{1}^{(2)}(0)$ if we choose the parameters of the system appropriately. In fig.\ref{fig5b} the smallest value of $g_{1}^{(2)}(0)$ is 0.063 and it is obtained for $\Delta_2 = 0.35 \omega_m$ and $E_2 = 0.05 \omega_m$. If we had cavity 1 decoupled from cavity 2, the smallest value of $g^{(2)}(0)$ would be 0.32. In fig.\ref{fig5c} we can see that the value of $g_{1}^{(2)}(0)$ is strongly dependent on the value of $\kappa_1$, making the condition $g \geq \kappa_1$ necessary to reach a significant reduction of $g_{1}^{(2)}(0)$. Among all figures, the smallest value for $g_{1}^{(2)}(0)$ was obtained in fig.\ref{fig5a}, $g_{1}^{(2)}(0) = 0.009$ for $\Delta_2 = 0.35\omega_m$ and $E_2 = 0.005 \omega_m$. These results reinforce that the system studied here is a potential candidate to a single-photon source \cite{stannigel}. 

In fig. \ref{fig6a} we analyze the relation between the minimum value achieved for $g_{1}^{(2)}(0)$ and $J$. We observe that, in the weak tunneling regime considered here, by increasing the value of $J$ it is possible to reach smaller values for $g_{1}^{(2)}(0)$. In the last section we conjectured that high enough values of $J$ could eventually compromise the sub-Poissonian character of the radiation field. Nonetheless, the results suggest that such high values of J would occur out of the weak tunneling regime considered here. In fig.\ref{fig6b} we plot the minimum value of $g_{1}^{(2)}(0)$ against g for different values of $E_2/E_1$ and $\overline{n}$. The continuous curves were obtained using the analytical approach presented in this paper; in which the mechanical oscillator is in contact with a heat bath at zero temperature ($\overline{n}=0$). The dashed curves were obtained using numerical simulations with the mechanical oscillator in contact with a heat bath with an occupation number $\overline{n}=1$. In the black curves (continuous and dashed) we have $J=0$, a situation which corresponds to the system considered in \cite{rabl}. In the blue curves (continuous and dashed) we have $J=0.05 \omega_m$ and $E_1 = E_2$, and in the red curves (continuous and dashed) we have $J=0.05 \omega_m$ and $E_2 = 5 E_1$. The peaks observed in all curves, as pointed out by Rabl, correspond to resonant transitions from a 1-photon state to a 2-photon state enabled by the absorption of phonons by the radiation field. We can observe, except in those peaks, a significant reduction of $min\{g_{1}^{(2)}(0)\}$, especially in the red curve. That behavior reinforces the results obtained in fig.\ref{fig5}. With respect to the dashed curves, a nonzero occupation number has significantly increased $min\{g_{1}^{(2)}(0)\}$, specially in the peaks, although the radiation still presents sub-Poissonian character. Particularly in the red curves, even with $\overline{n}=1$, it is still possible the find radiation with a very strong sub-Poissonian character. 

\section*{4. CONCLUSIONS}
In this work we have shown that by coupling two optical cavities optomecanically as well as by tunneling, it is possible to enhance the sub-Poissonian properties of one of the cavity fields. The results show that a more significant reduction of the second order correlation function $g_{1}^{(2)}(0)$ is possible only if the optomechanical coupling $g$ is larger than the decay rate $\kappa/2$ of the corresponding cavity. A better reduction of $g_{1}^{(2)}(0)$ can be obtained for certain values of the tunneling constant $J$. Our results also show that if the above conditions are satisfied, $g_{1}^{(2)}(0)$ can be significantly reduced by appropriately choosing parameters which are easily controlled in the laboratory, like $\Delta_2$ and $E_2$. Another important point is that even with $\overline{n}=1$, a significant reduction of $g_{1}^{(2)}(0)$ is possible. The results obtained reinforce the possibility of using similar systems as single photon sources.  

\begin{acknowledgments}
This work was supported by the São Paulo Research Foundation (FAPESP)(project No. 2012/10476-0), by the National Council for Scientific and Technological Development (CNPq), by the Optics and Photonics Research Center (CePOF) and the Brazilian National Institute for Science and Technology of Quantum Information (INCT-IQ).
\end{acknowledgments}


\begin{thebibliography}{99}
\bibitem{braginsky} V. B. Braginsky and A. B. Manukin, ``Ponderomotive effects of eletromagnetic radiation'', Sov. Phys. JETP {\bf 25}, 653 (1967).
\bibitem{braginsky2} V. B. Braginsky, A. B. Manukin and M. Y. Tikhonov, ``Investigation of dissipative ponderomotive effects of electromagnetic radiation'', Sov. Phys. JETP {\bf 31}, 829 (1970).
\bibitem{braginsky3} V. B. Braginsky, {\it Measurement of Weak Forces in Physics Experiments} (University of Chicago Press, 1977).
\bibitem{fabre} C. Fabre, M. Pinard, S. Bourzeix, A. Heidmann, E. Giacobino and S. Reynaud, ``Quantum-noise reduction using a cavity with a movable mirror'', Phys. Rev. A {\bf 49}, 1337 (1994).
\bibitem{mancini2} S. Mancini and P. Tombesi, ``Quantum noise reduction by radiation pressure'', Phys. Rev. A {\bf 49}, 4055 (1994).
\bibitem{jacobs} K. Jacobs, P. Tombesi, M. J. Collett and D. F. Walls, ``Quantum-nondemolition measurement of photon number using radiation pressure'', Phys. Rev. A {\bf 49}, 1961 (1994).
\bibitem{pinard} M. Pinard, C. Fabre and A. Heidmann, ``Quantum-nondemolition measurement of light by a piezoelectric crystal'', Phys. Rev. A {\bf 51}, 2443 (1995).
\bibitem{marquardt} F. Marquardt and S.M. Girvin, ``Optomechanics'', Physics {\bf 2}, 40 (2009).
\bibitem{mancini} S. Mancini, V. Giovannetti, D. Vitali and P. Tombesi, ``Entangling macroscopic oscillators exploiting radiation pressure'', Phys. Rev. Lett. {\bf 88}, 120401 (2002). 
\bibitem{vitali} D. Vitali, S. Gigan, A. Ferreira, H. R. Bohm, P. Tombesi, A. Guerreiro, V. Vedral, A. Zeilinger, and M. Aspelmeyer, ``Optomechanical entanglement between a movable mirror and a cavity field'', Phys. Rev. Lett. {\bf 98}, 030405 (2007).
\bibitem{arcizet} O. Arcizet, P.F. Cohadon, T. Briant, M. Pinard and A. Heidmann, ``Radiation-pressure cooling and optomechanical instability of a micromirror'', Nature {\bf 444}, 71 (2006).
\bibitem{gigan} S. Gigan, H. R. Böhm, M. Paternostro, F. Blaser, G. Langer, J. B. Hertzberg, K. C. Schwab, D. Bäuerle, M. Aspelmeyer and A. Zeilinger, ``Self-cooling of a micromirror by radiation pressure'', Nature {\bf 444}, 67 (2006).
\bibitem{schliesser} A. Schliesser, P. Del'Haye, N. Nooshi, K. J. Vahala and T. J. Kippenberg, ``Radiation pressure cooling of a micromechanical oscillator using dynamical backaction'', Phys. Rev. Lett. {\bf 97}, 243905 (2006).
\bibitem{thompson} J. D. Thompson, B. M. Zwickl, A. M. Jayich, Florian Marquardt, S. M. Girvin and J. G. E. Harris, ``Strong dispersive coupling of a high-finesse cavity to a micromechanical membrane'', Nature {\bf 452}, 72 (2008).
\bibitem{teufel} J. D. Teufel, J.W. Harlow, C. A. Regal, and K.W. Lehnert, ``Dynamical backaction of microwave fields on a nanomechanical oscillator'', Phys. Rev. Lett. 101, 197203 (2008).
\bibitem{rocheleau} T. Rocheleau, T. Ndukum, C. Macklin, J. B. Hertzberg, A. A. Clerk and K. C. Schwab, ``'Preparation and detection of a mechanical resonator near the ground state of motion', Nature {\bf 463}, 72 (2009).
\bibitem{teufel3} J. D. Teufel,	T. Donner, Dale Li, J. W. Harlow, M. S. Allman, K. Cicak,	A. J. Sirois,	J. D. Whittaker, K. W. Lehnert and R. W. Simmonds, ``Sideband cooling of micromechanical motion to the quantum ground state'', Nature {\bf 475}, 359 (2011).
\bibitem{groblacher} S. Gröblacher, K. Hammerer, M.R. Vanner and M. Aspelmeyer, ``Observation of strong coupling between a micromechanical resonator and an optical cavity field'', Nature {\bf 460}, 724 (2009).
\bibitem{teufel2} J.D.Teufel, Dale Li, M.S. Allman, K. Cicak, A.J. Sirois, J.D. Whittaker and R.W. Simmonds, ``Circuit cavity electromechanics in the strong-coupling regime'', Nature {\bf 471}, 204 (2011).
\bibitem{weis} S. Weis, R. Rivière, S. Deléglise, E. Gavartin, O. Arcizet, A. Schliesser, T. J. Kippenberg, ``Optomechanically Induced Transparency'', Science {\bf 330}, 1520 (2010).
\bibitem{safavi} A. H. Safavi-Naeini, T. P. Mayer Alegre, J. Chan, M. Eichenfield, M. Winger, Q. Lin, J. T. Hill, D. E. Chang	 and O. Painter, ``Electromagnetically induced transparency and slow light with optomechanics'', Nature {\bf 472}, 69 (2011).
\bibitem{bose} S. Bose, K. Jacobs and P. L. Knight, ``Preparation of nonclassical states in cavities with a moving mirror'', Phys. Rev. A, {\bf 56}, 4175 (1997).
\bibitem{mancini3} S. Mancini, V. I. Man'ko and P. Tombesi, ``Ponderomotive control of quantum macroscopic coherence'', Phys. Rev. A {\bf 55}, 3042 (1997).
\bibitem{rabl} P. Rabl, ``Photon blockade effect in optomechanical systems'', Phys. Rev. Lett. {\bf 107}, 063601 (2011).
\bibitem{kronwald} A. Kronwald, M. Ludwig, and F. Marquardt, ``Full photon statistics of a light beam transmitted through an optomechanical system'', Phys. Rev. A {\bf 87}, 013847 (2013).
\bibitem{stannigel} K. Stannigel, P. Komar, S. J. M. Habraken, S. D. Bennett, M. D. Lukin, P. Zoller and P. Rabl, ``Optomechanical quantum information processing with photons and phonons'', Phys. Rev. Lett. {\bf 109}, 013603 (2012).
\bibitem{qiu} L. Qiu, L. Gan, W. Ding, and Z-Y. Li, ``Single-photon generation by pulsed laser in optomechanical system via photon blockade effect'', J. Opt. Soc. Am. B {\bf 30}, 1683 (2013).
\bibitem{nunnenkamp} A. Nunnenkamp, K. Borkje and S.M. Girvin, ``Single-photon optomechanics'', Phys. Rev. Lett. {\bf 107}, 063602 (2011).
\bibitem{xu} X-W. Xu, Y-J. Li and Y-X. Liu, ``Photon-induced tunneling in optomechanical systems'', Phys. Rev. A {\bf 87}, 025803 (2013).
\bibitem{ludwig} M. Ludwig, A. H. Safavi-Naeini, O. Painter and F. Marquardt, ``Enhanced quantum nonlinearities in a two-mode optomechanical system'', Phys. Rev. Lett. {\bf 109}, 063601 (2012).
\bibitem{gupta} S. Gupta, K. L. Moore, K.W. Murch, and D. M. Stamper-Kurn, ``Cavity nonlinear optics at low photon numbers from collective atomic motion'', Phys. Rev. Lett. {\bf 99}, 213601 (2007).
\bibitem{brennecke} F. Brennecke, S. Ritter, T. Donner, and T. Esslinge, ``Cavity optomechanics with a Bose-Einstein condensate'', Science {\bf 322}, 235 (2008).
\bibitem{tan} S. M. Tan, ``A computational toolbox for quantum and atomic optics'', J. Opt. B {\bf 1}, 424 (1999).
\end{thebibliography}
\end{document}